\documentclass[fleqn,12pt,twoside]{article}
\usepackage[headings]{espcrc1}

\readRCS
$Id: espcrc1.tex,v 1.2 2004/02/24 11:22:11 spepping Exp $
\usepackage{graphicx}
\usepackage[figuresright]{rotating}

\newcommand{\AmS}{{\protect\the\textfont2
  A\kern-.1667em\lower.5ex\hbox{M}\kern-.125emS}}

\newcommand{\tr}{\mbox{Tr$\:$}} 

\title{Polyakov loop in different representations of
          SU(3) at finite temperature}

\author{S. Gupta,\address[mumbai]{Departement of theoretical Physics, Tata
    Institute for fundamental Research,\\ 
    Homi Bhabha Road, Mumbai 400005, India}
        K.~H\"ubner\address[bielefeld]{Fakult\"at f\"ur Physik, Universit\"at
        Bielefeld,\\ Universit\"atsstrasse 15, D-33615 Bielefeld, Germany}
              and
        O.~Kaczmarek\addressmark[bielefeld]
       }

\runtitle{Polyakov loop in different representations of
          SU(3) at finite temperature}

\begin{document}

\maketitle

\begin{abstract}
We investigate the Polyakov loop in different representations of
SU(3) in pure gauge at finite $T$.
We discuss Casimir scaling for
the Polyakov loop in the deconfined phase  and test and generalize the renormalization procedure
for the Polyakov loop from \cite{Kacz2002} to arbitrary representations.
In the confined phase we extract the renormalized adjoint Polyakov loop, which is finite in
the thermodynamic limit.
For our numerical calculations we used the tree-level improved
Symanzik action on $32^3\times 4,6,8$ lattices.
\end{abstract}

\section{The Polyakov loop in different representations}

We define the thermal Wilson line in representation $r$ 
\begin{equation}
  P_r(x)=\prod_{n=1}^{N_\tau} U^r_4(x+n\hat{t}),
\end{equation}
where $U_t$ are the temporal gauge links in representation $r$.
The Polyakov loop in  
representation $r$ is then 
\begin{equation}
  L_{r}(x) = \tr P_r(x),
\end{equation}
where $\tr 1=1$.
We obtain $L_r(x)$ from $L_3(x)$
by the aid of the character property of the direct product representation  
$\chi_{p\otimes q}(g)=\chi_p(g)\chi_q(g)$ (see table I in \cite{Bali00}).
The expectation value of the Polyakov loop is then obtained by
\begin{equation}
  \langle L_{r}\rangle = 
  \left\langle\frac1{V_3}\sum_x L_{r}(x)\right\rangle,
\end{equation}
where $V_3$ is the spatial volume of the lattice.
We have calculated the expectation values of the Polyakov loop in
  higher representations up to $r=15a$ in pure gauge theory.

\section{Renormalization and Casimir scaling}

\begin{figure}[t]
\center{
  \includegraphics[width=7.5cm]{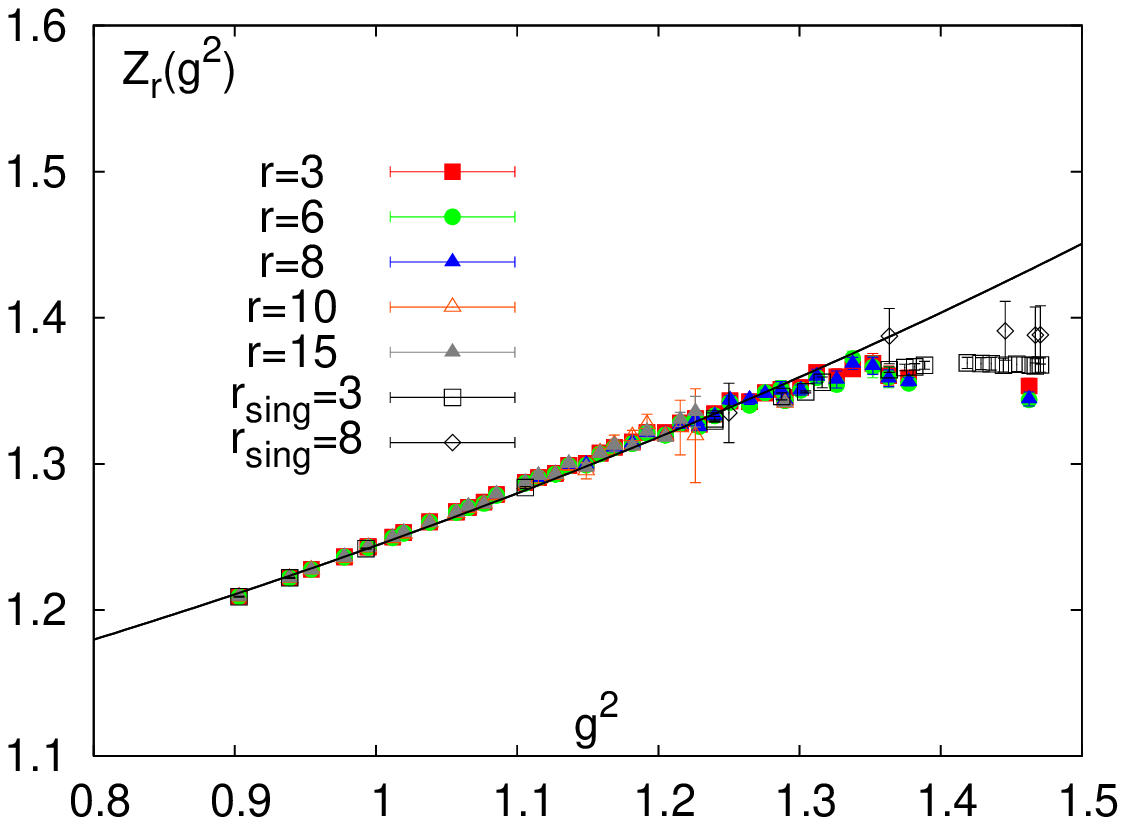}
  \includegraphics[width=7.5cm]{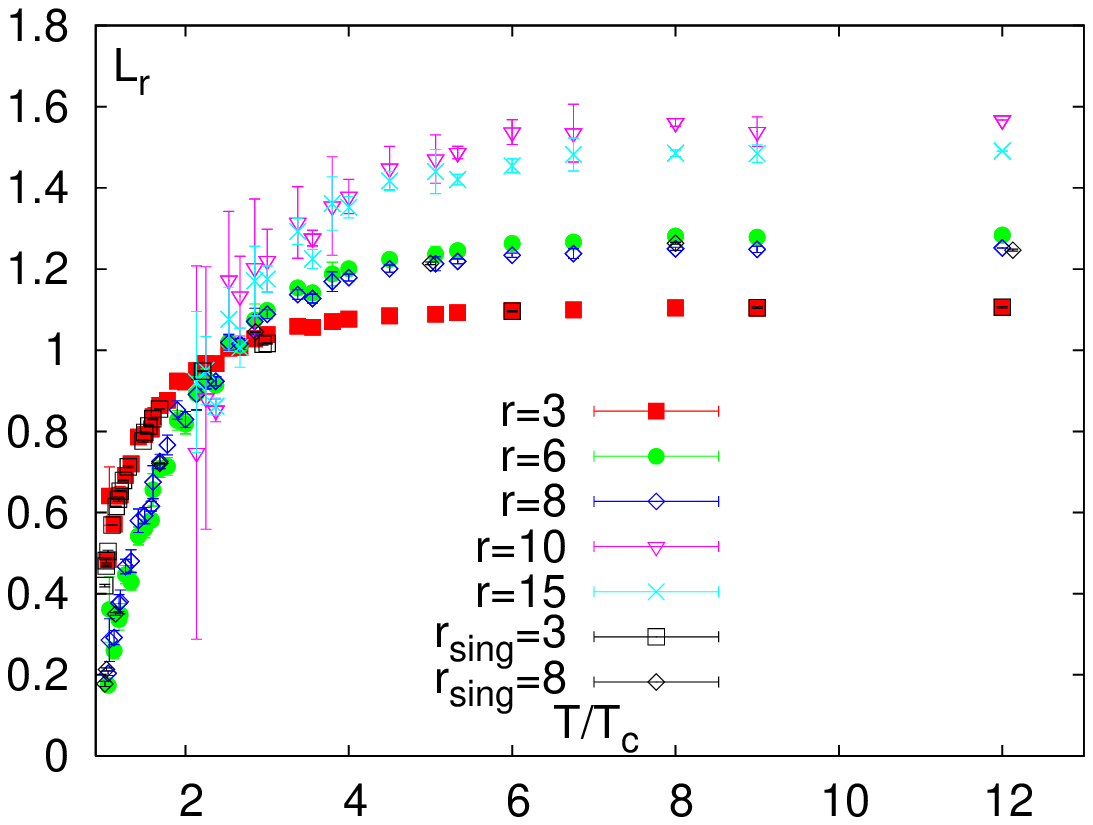}
  \caption{\label{fig:all_renorm}Left: renormalisation constants $Z_r(g^2)$ for
  $r\le 15$ obtained by the $N_\tau$-variation method and for $r=3,8$ from the
  matching of the $Q\bar Q$-singlet free energies to the $T=0$
  potential. Right: renormalised Polyakov loop.
}
}
\end{figure}

Realizing that the renormalization constants only depend on the bare coupling \cite{Brandt:1981kf},
$Z_r(g^2)$, the renormalized Polyakov loop can be obtained from the bare loops
at different $N_\tau$. As any renormalization procedure is fixed up to a
multiplicative constant , i.~e.~$L_r^c=L_re^{-c/2T}$, we fix the value of the
renormalized Polyakov loop at the highest temperature ($12T_c$) to the value
obtained by the renormalization procedure used in \cite{Kacz2002}. 
We obtain the corresponding $Z_r(g^2)$ for the different $N_\tau$ and
can use them to renormalize the Polyakov loop at a different $T$ by
varying $N_\tau$. At this new value of $T$, all values of $L_r^{(b)}$
for different $N_\tau$ can be renormalized to this value and
determine a new series of $Z_r(g^2)$ at different couplings
corresponding to this $T$ at the various $N_\tau$. The same method
can now be performed for all representations. We only have to fix the
value at the highest temperature, where we assume Casimir scaling,
\begin{equation}
  \label{eq:casimir_scaling_l_r}
  L_r(12T_c)\equiv (L_3(12T_c))^{d_D},
\end{equation}
where $d_D=C_2(r)/C_2(3)$. In perturbation theory the renormalization constants
are related by Casimir scaling (at least up to two loop order ($\alpha_s^3$)
\cite{Schroder99}). Therefore we define them through 
\begin{equation}
  \label{eq:pol_renorm_scheme}
  L_r=(Z_r(g^2))^{N_\tau d_D}L_r^{(b)}.
\end{equation}
The results for $Z_r(g^2)$ and $L_r(T)$ are shown in 
fig.~\ref{fig:all_renorm}. We observe that the
renormalization constants obtained independently in this way agree surprisingly
well in the whole coupling range analyzed here. 
Furthermore they also agree with the renormalization constants obtained for
$r=3$ and $r=8$ with the renormalization procedure outlined in \cite{Kacz2002},
that defines the renormalized fundamental Polyakov loop 
by renormalizing the singlet free energy at small distances to the
(perturbative) zero temperature potential. 
The solid line in fig.~\ref{fig:all_renorm}(left) shows the result of a best-fit
analysis to the perturbative inspired ansatz 
\begin{equation}
Z_r(g^2)=\exp\left(g^2\frac{8}{3}Q^{(2)}+g^4Q^{(4)}\right),
\end{equation}
where we find $Q^{(2)}=0.0576(25)$ and $Q^{(4)}=0.0639(68)$.

If Casimir scaling is realized, the renormalization constants for different
representations should agree, i.~e.~
\begin{equation}
  \label{eq:pol_renorm}
  Z_r(g^2)=Z_{3}(g^2).
\end{equation}
As we have observed that the renormalization constants are related by Casimir
scaling, we can now analyze if Casimir scaling
also holds for the Polyakov loops. Through (\ref{eq:pol_renorm_scheme}) it is
equivalent to analyze this for the bare or renormalized loops. 
If  Casimir scaling for the bare Polyakov loops is realized,
\begin{equation}
  \label{eq:casimir_pol}
  \left\langle L_r^{(b)}\right\rangle=\left\langle
    L_{3}^{(b)}\right\rangle^{d_r},
\end{equation}
we expect the quantity $\langle L_{r}^{(b)}\rangle^{1/d_r}$ to be independent
of the representation $r$.
In fig.~\ref{fig:all_pol}(left) we show the results of our calculations.
We observe all data to collapse onto a single curve for temperatures down close
to the critical temperature,
confirming (\ref{eq:casimir_pol}) to be valid within errors for 
representations $r=3,6,8,10,15$ above $T_c$. 

\section{The adjoint Polyakov loop in the confined phase}

\begin{figure}[t]
\center{
  \includegraphics[width=7.5cm]{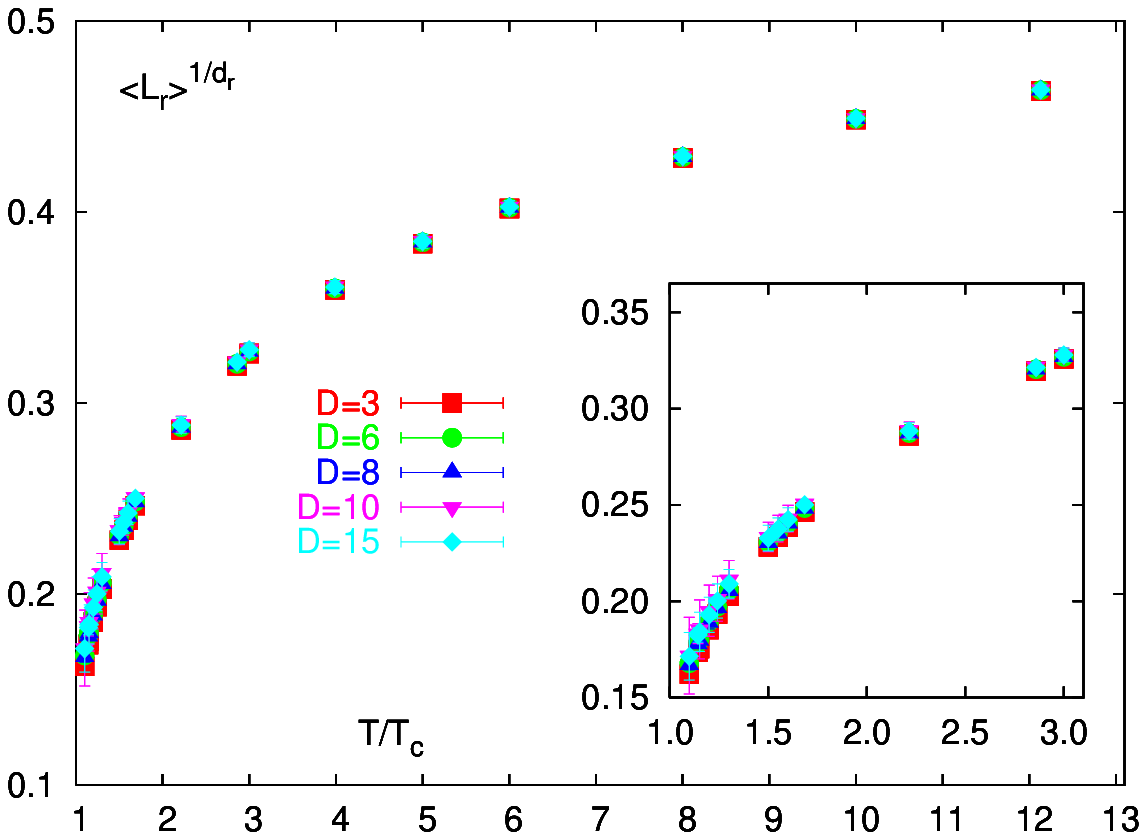}
  \includegraphics[width=7.5cm]{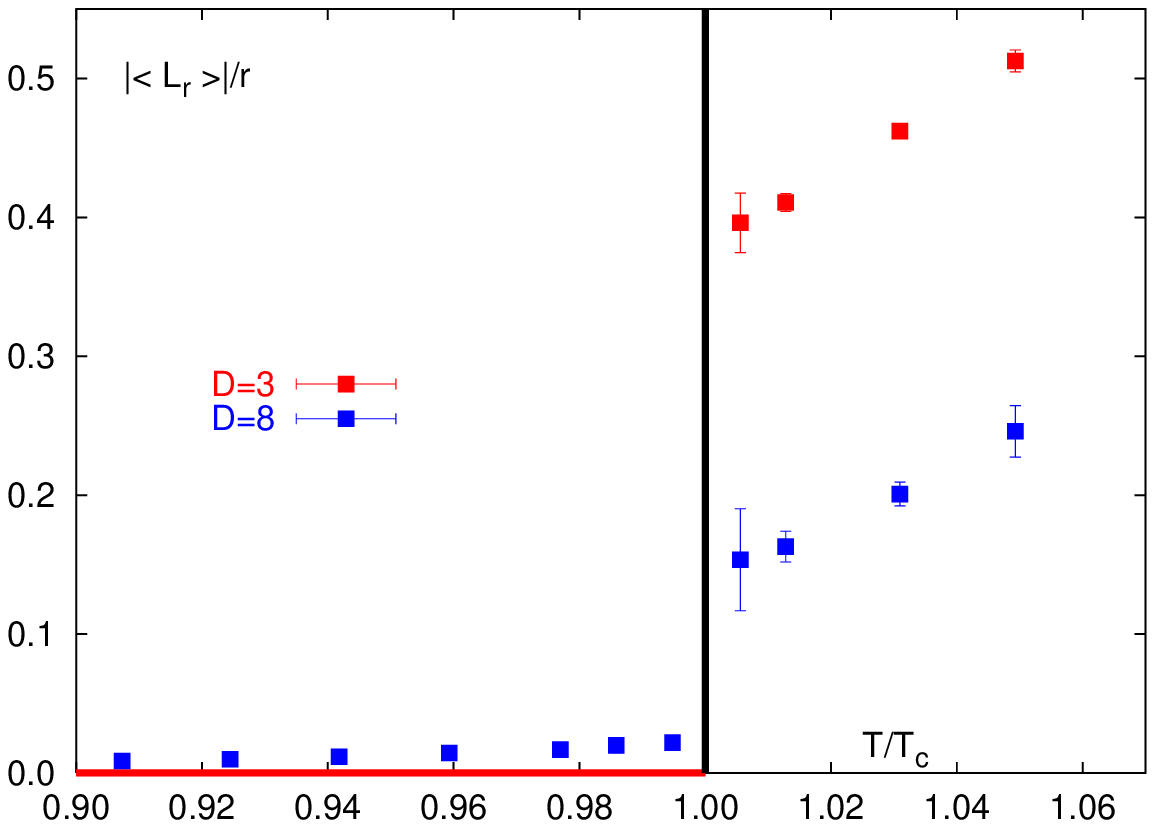}
  \caption{\label{fig:all_pol}
  Left: bare Polyakov loops for $r\le 15$ with cubic spline above
  $T_c$. Right: renormalized fundamental and adjoint Polyakov loop for
  temperatures around $T_c$.}
}
\end{figure}

While the Polyakov loop in the fundamental representation vanishes in the
thermodynamic limit in the confined phase, 
for all triality-zero representations (r=8,10,27), one expects to see
string breaking below $T_c$ also in pure gauge theory
\cite{Bali00}, and hence a non-vanishing Polyakov loop in the infinite
  volume limit.

We have analyzed the volume dependence of the adjoint Polyakov loops and found
that for $N_\sigma=32$ the infinite volume limit is already reached.
We can therefore use the $N_\sigma=32$ data 
to obtain the infinite volume, renormalized adjoint Polyakov
  loop below $T_c$. 
Fig.~\ref{fig:all_pol}(right) shows $\left\langle L_8\right\rangle$ around
$T_c$. While the adjoint Polyakov loop is small but non-zero for all
temperatures analyzed here, the fundamental renormalized Polyakov
loop is zero below $T_c$.

For the other triality-zero representations ($r=10,27$), we cannot
give the infinite volume limit below $T_c$, since the corresponding
$\left\langle L_r\right\rangle$ still
show volume dependences on the analyzed lattices.

\section{Conclusion and Outlook}
 
We have investigated the Polyakov loop and free energies 
of static quark-antiquarks in different representations of SU(3) in pure gauge at finite $T$.
Above the crititcal temperature the Polyakov loop in different
representations shows Casimir scaling down close to $T_c$.
Moreover, we were able to test the renormalization scheme
proposed in \cite{Kacz2002} for the fundamental Polyakov loops and could
generalize it to arbitrary representations. Furthermore we also demonstrate
that the renormalization constants only depend on the bare coupling, at
least up to $g^2\approx 1.3$.
Below the critical temperature we were
able to show, that the adjoint
Polyakov loop is finite in the infinite volume limit and extract its
infinite volume renormalized value. For higher triality-zero
representations, no definite answer could be given within this work.
We note that the renormalization procedure discussed here is based solely on
gauge independent  quantities and that the agreement of the renormalized
Polyakov loops and the renormalization constants with the procedure outlined in
\cite{Kacz2002} suggests that also in that procedure no gauge dependence exists.
For discussion on potential gauge dependencies see \cite{Philipsen:2002az,Jahn:2004qr}.
A different renormalization procedure was proposed in \cite{Dumitru:2003hp}.

\section*{Acknowledgement}

We wish to thank J.~Engels, F.~Karsch, R.~D.~Pisarski and F.~Zantow for fruitful discussions. K.~H.~was supported by DFG
under grant GRK 881/1.

\end{document}